\documentclass[reprint,twocolumn,superscriptaddress,aps]{revtex4-1}
\usepackage{graphicx}
\usepackage{amsmath}
\usepackage{latexsym}
\usepackage{amsfonts}
\usepackage{amssymb}
\usepackage{array}
\usepackage{epsfig}
\usepackage[colorlinks=true,linkcolor=blue,urlcolor=blue,citecolor=blue]{hyperref}
\usepackage{color}
\usepackage{hyperref}
\usepackage{braket}
\usepackage{ulem}
\setcitestyle{numbers,square,citesep={,\kern-2pt}}

\begin{document}

\title{Schmidt gap in random spin chains}
\author{Giacomo Torlai}
\thanks{These two authors contributed equally.}
\affiliation{Department of Physics and Astronomy, University of Waterloo, Ontario N2L 3G1, Canada}
\affiliation{Perimeter Institute for Theoretical Physics, Waterloo, Ontario N2L 2Y5, Canada}
\author{Kenneth D. McAlpine}
\thanks{These two authors contributed equally.}
\affiliation{Centre  for  Theoretical  Atomic,  Molecular  and  Optical  Physics, Queen's  University  Belfast,  Belfast  BT7 1NN,  United  Kingdom}
\author{Gabriele De Chiara}
\affiliation{Centre  for  Theoretical  Atomic,  Molecular  and  Optical  Physics, Queen's  University  Belfast,  Belfast  BT7 1NN,  United  Kingdom}

\begin{abstract}
We numerically investigate the low-lying entanglement spectrum of the ground state of random one-dimensional spin chains obtained after partition of the chain into two equal halves. We consider two paradigmatic models: the spin-1/2 random transverse field Ising model, solved exactly, and the spin-1 random Heisenberg model, simulated using the density matrix renormalization group. In both cases we analyze the mean Schmidt gap, defined as the difference between the two largest eigenvalues of the reduced density matrix of one of the two partitions, averaged over many disorder realizations. We find that the Schmidt gap detects the critical point very well and scales with universal critical exponents.
\end{abstract}

\maketitle

\section{Introduction}
The entanglement spectrum (ES) \cite{LiHaldane}, the set of eigenvalues of the reduced density matrix of a quantum many-body state, has now become a standard fingerprint that reveals much more information on a state compared to measures of bipartite entanglement, such as the von Neumann entropy and the negativity (see Refs.~\cite{LaflorencieReview} and \cite{DeChiaraSanperaROPP} for recent comprehensive reviews). 

Originally introduced to study the transition to a topologically ordered state in the quantum Hall effect \cite{LiHaldane}, ES has been used for the characterization of spin chains and other one-dimensional (1D) models in real and momentum space \cite{Xu2008,PollmannMoore2010,Thomale2010,Poilblanc2010,DengSantos2011,AlbaPRL2012,LauchliPRB2012,KorepinEPL2012,PouranvariPRB2013,LundgrenPRB2016}. The distribution of the Schmidt eigenvalues in the middle of the spectrum has been studied by means of conformal field theory \cite{CalabreseLefevre}. The study of the structure of the low-lying part of the ES in 1D models also reveals the Luttinger parameter \cite{Lauchli2013,LaflorencieRachel2014}. In 2D systems the situation is somewhat less clear and the universality of the ES has been challenged~\cite{ChandranPRL2014}.

The Schmidt gap, the difference of the two largest Schmidt eigenvalues of the ES, originally introduced in \cite{PollmannPRB2010} and \cite{DeChiaraPRB2011} was shown to scale according to universal critical exponents in \cite{DeChiara12,Lepori13,GiampaoloPRB2013,Bayat2014,GallemiPRA2016}. It was further employed in the characterization of 2D spin models in a region close to a topological spin liquid \cite{Moreno2014, Mandal16}. The time evolution of the Schmidt gap was analyzed in \cite{Torlai2014,CanoviPRB2014,HuPRB2015}  for the dynamics after a quantum quench in homogeneous systems and in \cite{Gray18} for a quench to a many-body localized Hamiltonian. Whether or not the Schmidt gap can be applied as an instrument to detect criticality in random models is still an open question.

The effect of randomness in spin models, whether introduced via disorder in coupling constants or through some random external field, has become an area of significant interest since the early studies on the random transverse field Ising model \cite{Fisher1992,Fisher1995,Young96}. Randomness has been shown to modify the characteristics of phase transitions \cite{Bunder99,McAlpine2017}, as well as transition a spin system from one universality class to another \cite{Lajko05,Laflorencie05} and is integral to the emergence of interesting phases such as the Griffiths and random singlet phases (RSPs) \cite{Griffiths69,Saguia02,Quito15}. 
Recently a lot of attention has been devoted to the mechanism of many-body localisation in 1D and 2D systems~\cite{Nandkishore2015,Altman2015,Abanin2017,Alet2018}.
While these random models have usually been investigated using corresponding order parameters \cite{Bergkvist02,Lajko05} and entanglement entropy \cite{Refael04,Saguia07,Ferenc12,Ruggiero16}, a few works have analyzed numerically the entanglement spectrum of the ground and excited states of random spin chains \cite{Fagotti11,PouranvariPRB2013,Yang15,Pouranvari15}.

In this paper, we study the Schmidt gap of the ground state of random spin-1/2 and spin-1 chains. We show for both models that the closing of the Schmidt gap, averaged over the disorder distribution, signals the occurrence of a quantum phase transition. Moreover, we are able to observe universal scaling of the Schmidt gap with critical exponents.

\section{Random transverse-field Ising model}
\label{sec:ising}
We consider $L$ spin-$\frac{1}{2}$ arranged in a chain with open boundary conditions and Hamiltonian
\begin{equation}
H=-\sum_i J_{i}\sigma^x_i \sigma^x_{i+1}-\sum_i h_i\sigma_i^z\:.
\label{randomIsingHamiltonian}
\end{equation}
The couplings $J_i$ of the Ising interaction and the transverse magnetic fields $h_i$ are independent random variables drawn from the distributions $\pi(J)dJ$ and $\rho(h)dh$, which can be gauged to be positive. In the following, we consider the distributions $\pi(J)$ and $\rho(h)$ to be uniform in the intervals $J\in[0,1]$ and $h\in[0,h^{\text{max}}]$, respectively, and 0 otherwise. This choice reduces the Hamiltonian parameters to only one variable, $h^{\text{max}}$. 

The physics underlying the ground state of this Hamiltonian is closely related to the finite-temperature behavior of a 2D classical Ising model with quenched randomness correlated along one direction~\cite{McCoy1968,McCoy1969}. The quantum Hamiltonian in Eq.~\eqref{randomIsingHamiltonian} is recovered by taking the continuum limit of the classical model, and it was first investigated with transfer matrix methods by Shankar and Murthy~\cite{Shankar1987}. In particular, by a simple duality argument, it can be shown that a quantum phase transition takes place when the two distributions $\pi(J)$ and $\rho(h)$ are identical. By defining the magnetic-field parameter $\Delta_h=[\log h]_D$ (where $[\:\cdot\:]_D$ is the disorder average), the quantum critical point is found at $\Delta_c=[\log J]_D$, corresponding to $h^{\text{max}}_c=1$ for our choice of distributions. The phase diagram features a paramagnetic phase ($\Delta_h>\Delta_c$) and a ferromagnetic phase ($\Delta_h<\Delta_c$) with non-zero spontaneous magnetization $m_x=[\sum_i \langle\sigma_i^x\rangle]_D\ne0$, where $\langle\cdot\rangle$ denotes the ground-state average.

The magnetic properties, both at criticality and off criticality, can be derived following a renormalization group approach~\cite{Fisher1992}, where the short-wavelength modes are cut off from the system by targeting the strongest coupling $\Omega=\text{max}\{J_i,h_i\}$. In practice, via perturbation theory, the  excited states of the subspace for the local Hamiltonian describing the degrees of freedom connected to $\Omega$ are eliminated, leading to a new effective Hamiltonian with a lower energy scale $\Omega$ at each step. This iterative procedure allows the estimation of the correlation function and the various critical exponents~\cite{Fisher1995}. In particular, the behavior of the typical correlation function $C(r)=\sum_i\langle\sigma_i^x\sigma_{i+r}^x\rangle$ is found to be very different from that of the average correlation function $[C(r)]_D=[\sum_i\langle\sigma_i^x\sigma_{i+r}^x\rangle]_D$. At criticality, the typical correlation decays as $C(r)\sim\text{exp}(-\sqrt{r})$ while the average correlation follows a power-law decay $[C(r)]_D\sim1/r^{2-\phi}$, where $\phi=(1+\sqrt{5})/2$ is the golden mean. On the other hand, in the paramagnetic phase, both correlation functions decay exponentially with correlation length $\xi\sim(h^{\text{max}}-h^{\text{max}}_c)^{-\nu}$. The critical exponent $\nu$ differs for the two cases, with $\nu=1$ and $\nu=2$ for the typical and average correlation function, respectively. The spontaneous magnetization in the ferromagnetic phase is $m_x(h^{\text{max}})\sim(h^{\text{max}}_c-h^{\text{max}})^\beta$ with critical exponent $\beta=2-\phi=(3-\sqrt{5})/2\simeq 0.381$.
\begin{figure}[t]
\noindent \centering{}
\includegraphics[trim=6.5cm 2.5cm 7.75cm 0.5cm, clip=true, width=0.95\columnwidth]{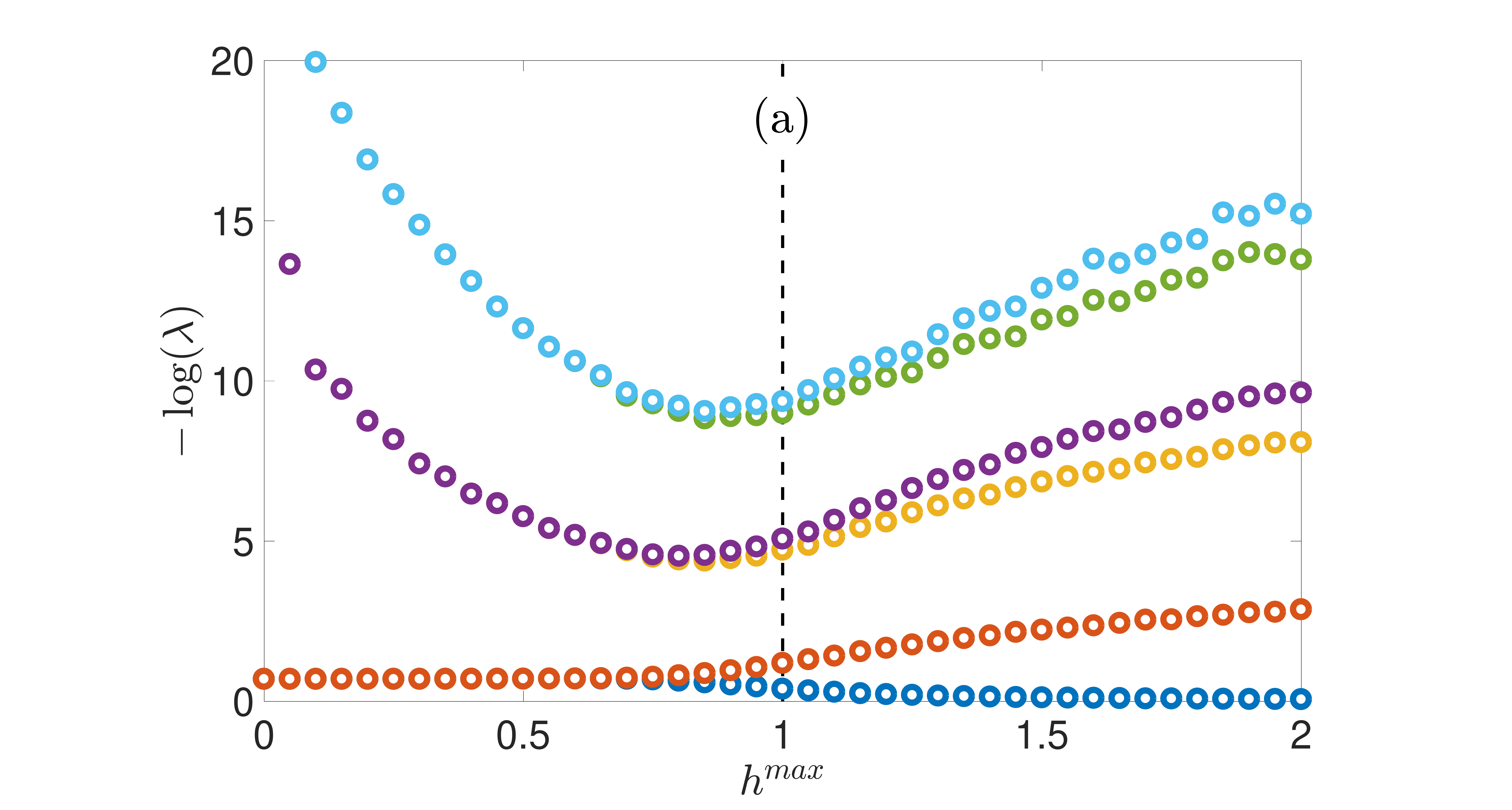}\\[-7.5pt]
\includegraphics[trim=6.25cm 0.45cm 8.1cm 1.5cm, clip=true, width=0.95\columnwidth]{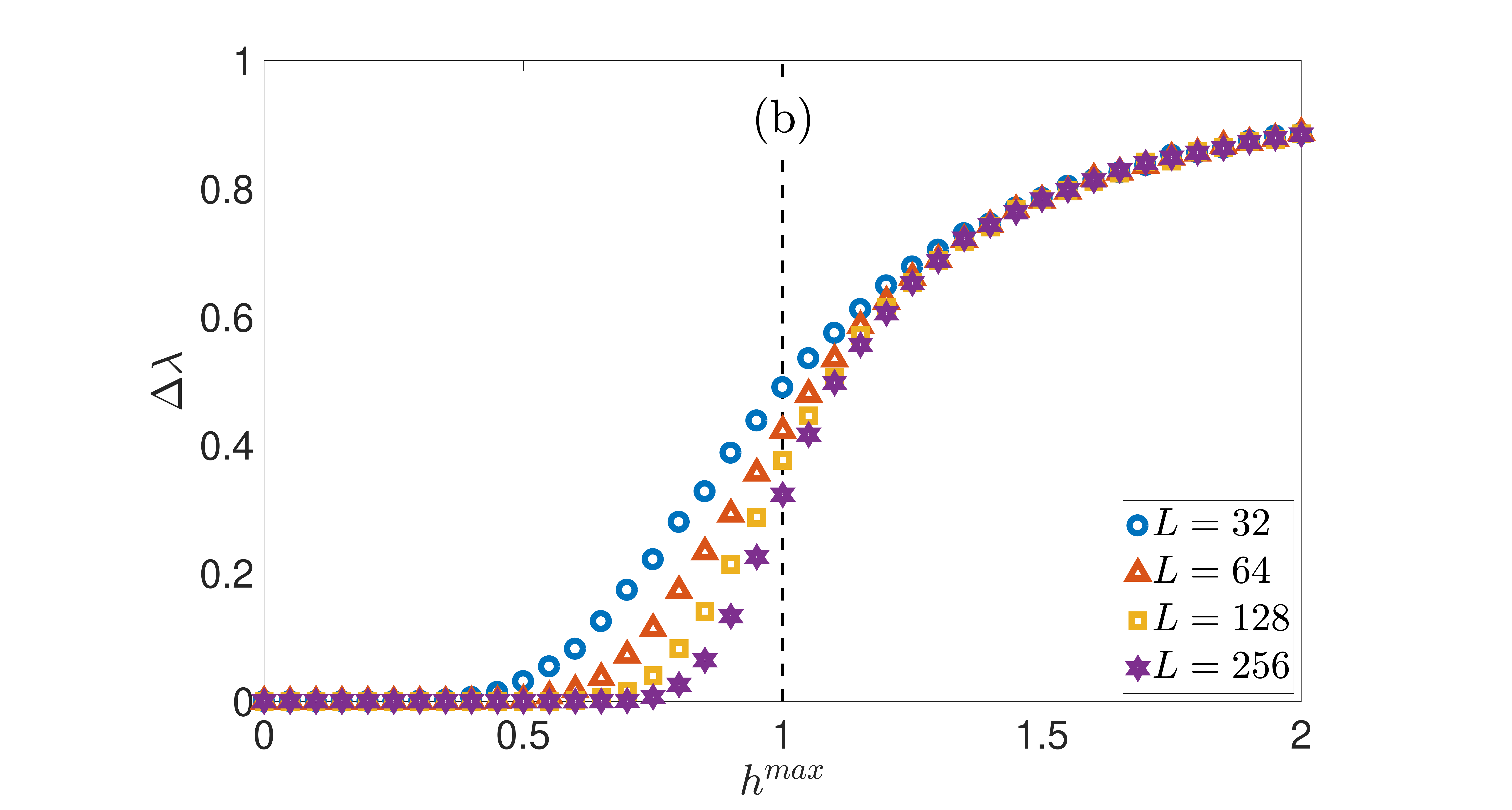}
\caption{Entanglement structure of the random transverse-field Ising chain, for a system size of $L=128$. (a) Largest eigenvalues $\lambda_i$ in the ES plotted versus the upper bound $h^{\text{max}}$ of the magnetic-field distribution $\rho(h)$. (b) Schmidt gap $\Delta\lambda$ as a function of $h^{\text{max}}$ for different systems sizes. Each data point was computed as the average over $10^4$ realizations of disorder.  The dashed line represents the expected critical value of $h^{\text{max}}$ at which the phase transition occurs.}
\label{fig:ising} 
\end{figure}

The entanglement structure for the random transverse-field Ising model can be calculated exactly by first mapping the spin degrees of freedom into a system of non interacting fermions using the Jordan-Wigner transformation~\cite{Lieb1961}. Within this representation, the reduced density matrix for a subsystem $S$ is simply given by $\rho_S=Z^{-1}\text{exp}(-K)$, where $K$ is called the entanglement Hamiltonian and $Z$ is the normalization constant~\cite{Peschel2009}. Given the correlation matrices $C=\langle c^\dagger c\rangle$, $F=\langle c^\dagger c^\dagger\rangle$, with $(c^\dagger,c)$ the fermionic creation and annihilation operators, the eigenvalues $\epsilon_k$ of $K$ can be calculated from the matrix $M=2C-I-2F$, where $I$ is the identity matrix, by singular value decomposition~\cite{Peschel2003}. We can easily calculate the ES $\{\lambda_i\}$ directly from the full spectrum $\{\epsilon_k\}$ of the entanglement Hamiltonian following the approach explained in Ref.~~\cite{Peschel2003}. In all calculations we partition the chain into two equal halves.

In Fig.~\ref{fig:ising} we show the entanglement properties as a function of the upper bound $h^{\text{max}}$ of the magnetic-field distribution $\rho(h)$. In Fig.~\ref{fig:ising}(a) we plot the six largest  eigenvalues of the ES for a system comprising $L=128$ spins. Each data point is obtained by averaging the eigenvalue over $10^4$ realizations of disorder. Analogously to the homogeneous case~\cite{DeChiara12}, in the ferromagnetic phase the ES is characterized by doubly degenerate multiplets, as a consequence of the unbroken $Z_2$ symmetry. The doublets are lifted at and beyond the quantum critical point $h_c^{\text{max}}=1$. 

We now study the properties of the Schmidt gap $\Delta\lambda=\lambda_1-\lambda_2$, where $\lambda_1$ and $\lambda_2$ are the two largest Schmidt eigenvalues. As shown in Fig.~\ref{fig:ising}(b), for small fields $h^{\text{max}}<1$, $\Delta\lambda\sim 0$ as a consequence of the existence of the doublets in the ferromagnetic phase. But for larger fields corresponding to the paramagnetic phase, $\Delta\lambda$ grows and eventually saturates to 1 for infinite magnetic field (all the spins are aligned along the magnetic field and the state is a product state with a single Schmidt eigenvalue). The behavior of the Schmidt gap in the random Ising chain is similar to its behavior in the corresponding homogeneous Ising chain \cite{DeChiara12}. Therefore it is quite intriguing to check whether the critical scaling of the Schmidt gap can be observed also in the random model.

To this end we assume a finite-size scaling ansatz for $\Delta\lambda$ which is normally employed for order parameters \cite{Fisher72,DeChiara12}:
\begin{equation}
\label{eq:fss}
Q(L,h^{\text{max}})\simeq L^{-\beta_Q/\nu}f_Q(|h^{\text{max}}_c-h^{\text{max}}|L^{1/\nu})
\end{equation}
where $Q$ is the order parameter under investigation and $\beta_Q$ the corresponding critical exponent.

\begin{figure}[t]
\noindent \centering{}
\includegraphics[trim=6.5cm 0.25cm 9cm 1.5cm, clip=true, width=0.95\columnwidth]{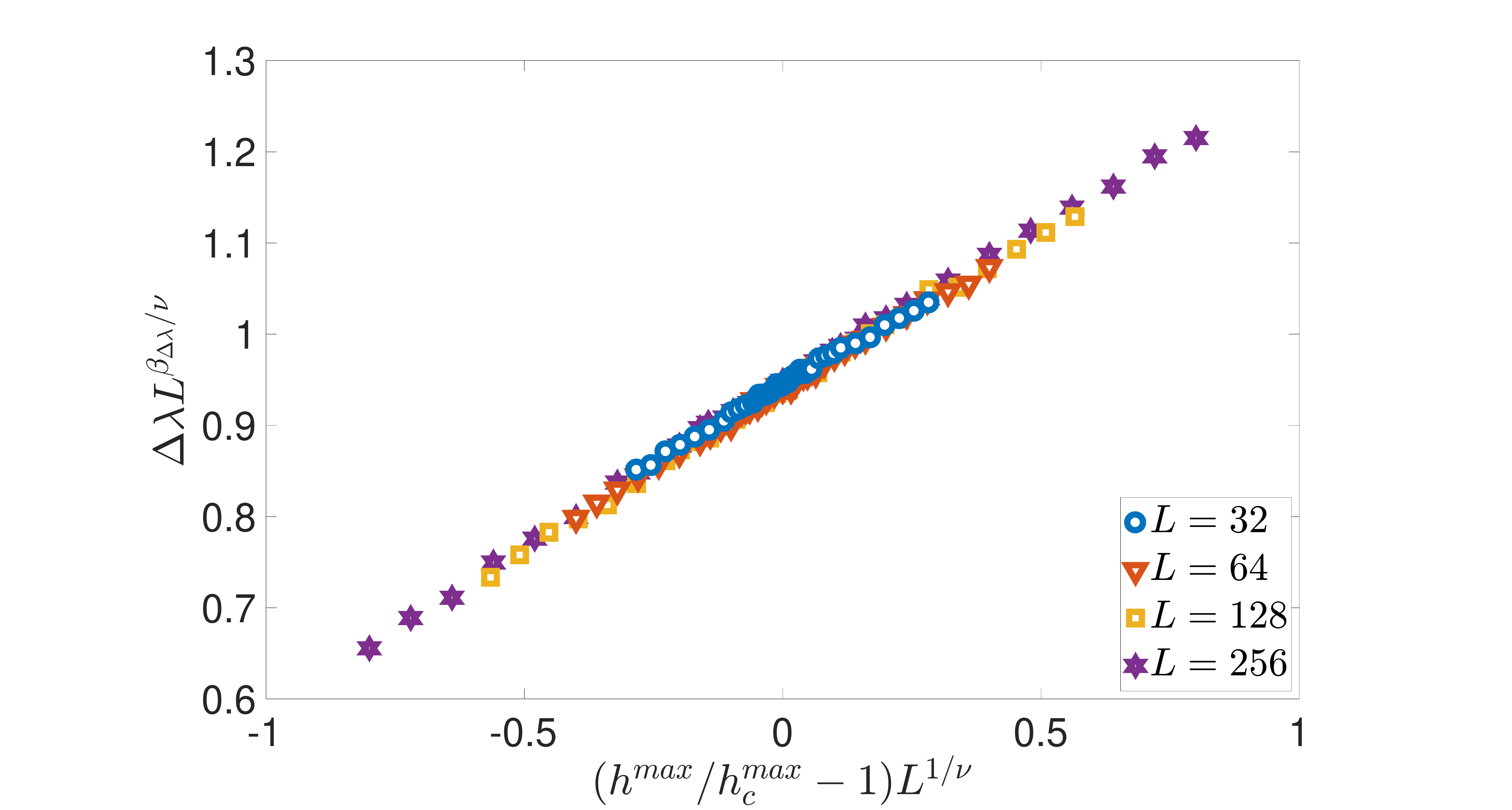}
\caption{Collapse plot for the Schmidt gap $\Delta\lambda$ as a function of $h^{\text{max}}$ for different system sizes, $L=32$, 64, 128, and 256, obtained averaging over $3\times 10^4$ realizations of disorder using Eq.~\eqref{eq:fss}. The critical exponent $\beta_{\Delta\lambda}=0.39 \pm 0.01$ is obtained from a fit with fixed parameters $h^{\text{max}}=1.0$ and $\nu=2.0$.}
\label{fig:isingcollapse} 
\end{figure}

 Using ansatz \eqref{eq:fss}, in Fig. \ref{fig:isingcollapse}, we performed a fit with system sizes $L=32$, 64, 128, and 256 around the critical point, where each point was averaged over $3\times10^4$  realizations of disorder. We observe the best data collapse with an estimated critical exponent $\beta_{\Delta\lambda}=0.39 \pm 0.01$ compatible (within statistical and finite-size error) with the order parameter critical exponent $\beta$. In the fit we fixed the correlation length critical exponent $\nu=2$.
 
  This numerical evidence clearly demonstrates that even for the random model, as for the homogeneous case, the low-lying structure of the ES is universal and determined by universal critical exponents.

\section{Random spin-1 Heisenberg chain}

In this section, we turn our investigation to the spin-1 random antiferromagnetic Heisenberg chain, defined by the Hamiltonian \cite{Hyman97,Monthus97,Lajko05},
\begin{equation}\label{Hamiltonian}
H=\sum_iJ_i\mathbf{S}_i\cdot\mathbf{S}_{i+1}
\end{equation}
where ${\mathbf{S}_i = (S_{xi},S_{yi},S_{zi})}$ are the $i$th-site angular momentum operators and $J_i$ are positive couplings.
For $J_i=J>0$, i.e., a homogeneous chain, the ground state is in the gapped Haldane phase, characterized by the absence of any local order, a nonzero string order, which we introduce later, and an evenly degenerate ES. 

We introduce disorder in the model by choosing:
\begin{equation}
J_i=\zeta_i^\delta,
\end{equation}
where $\delta$ controls the strength of disorder and $\zeta_i$ is a random variable distributed uniformly between 0 and 1. The probability distribution of $J_i$ is:
\begin{equation}
\label{eq:powerdist}
\pi_\delta(J)=\delta^{-1}J^{-1+1/\delta}.
\end{equation}

The gapped Haldane is stable for ${J_{\rm min}/J_{\rm max}>0.6}$, where $J_{\rm min}$ and $J_{\rm max}$ are the smallest and largest couplings, respectively.
For the power-law distribution, \eqref{eq:powerdist}, and for ${\delta>0}$, we have $J_{\rm min}/J_{\rm max}=0$, and the gapped Haldane phase is immediately destroyed for an infinitesimal amount of this type of disorder.
However, for small $\delta$ the system enters the so-called gapless Haldane phase, a type of Griffith phase with closed Haldane gap, but exhibiting the hidden topological order characteristic of the gapped phase \cite{Hyman97,Quito15}.
For very strong disorder $\delta\gg 1$, the ground state is in the random singlet phase, which is a gapless phase consisting of pairs of spins in singlets spanning arbitrarily long distances \cite{Ma79,Bhatt82,Fisher94,Monthus97}.

The phase diagram for the spin-1 random antiferromagnetic Heisenberg chain when using a power-law disorder distribution is the following \cite{Yang00,Carlon04,Lajko05}: gapped Haldane at zero disorder (${\delta=0}$), gapless Haldane (Griffiths) at ${0<\delta<1}$, and, finally, RSP at ${\delta\geq1}$.
This power-law distribution for the disorder is required in order to cross the phase transition between the Haldane and the RSPs \cite{Yang00,Lajko05}. The critical point at which this phase transition takes place is known to be approximately ${\delta_c=1}$. A box-like disorder distribution is only able to reach a disorder distribution equivalent to that of ${\delta=1}$ and, thus, is unable to cross the quantum phase transition to the RSP \cite{Hida99}.

The results reported in this section of the paper have been obtained using finite-size density matrix renormalization-group (DMRG) calculations with open boundary conditions~\cite{DMRG1,DMRG2}; between 2000 and 2500 random realizations were used. In an attempt to reduce issues in the calculations relating to degeneracy of the ground state, a staggered magnetic field of magnitude ${2.5\times10^{-3}}$ was placed on the first two and last two spins. Due to the spin chain having zero spontaneous magnetization for all values of $\delta$, we project over the total angular momentum ${M_z = \sum_iS_{zi} = 0}$. In the DMRG calculations 100 states were kept during the renormalization process, resulting in a maximum discarded weight of $10^{-6}$. We remark that an alternative method to deal with random spin chains is to employ a quantum parallel method in which disorder is simulated by means of auxiliary sites coupled to the physical sites~\cite{ParedesPRL2005}. However, within this method the calculation of the entanglement spectrum for each disorder realization would not be efficient.

\begin{figure*}[t] 
\centering
\includegraphics[trim=5.5cm 0.45cm 9.5cm 1.5cm, clip=true, width=0.65\columnwidth]{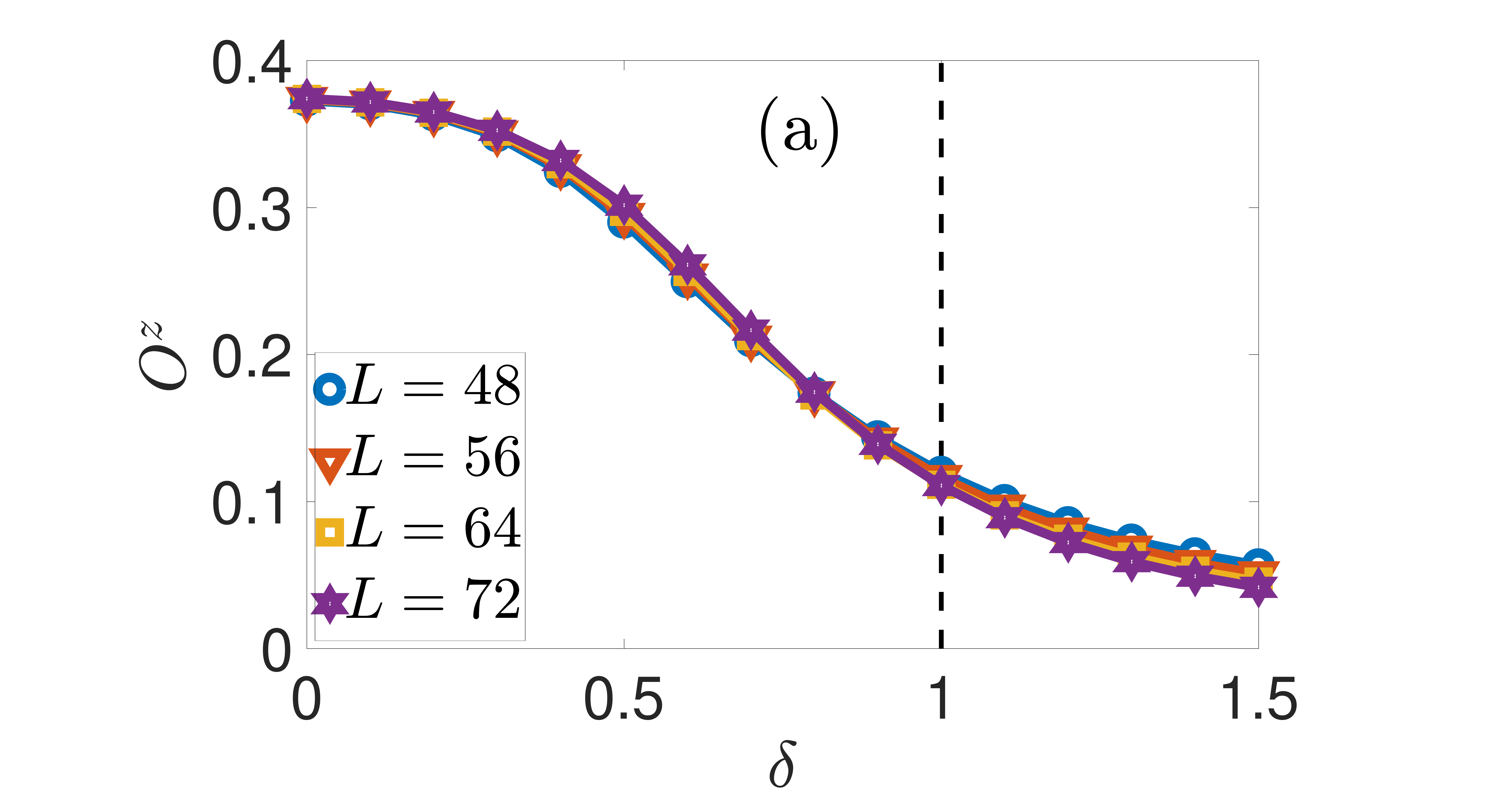}
\includegraphics[trim=5.5cm 0.45cm 9.5cm 1.5cm, clip=true, width=0.65\columnwidth]{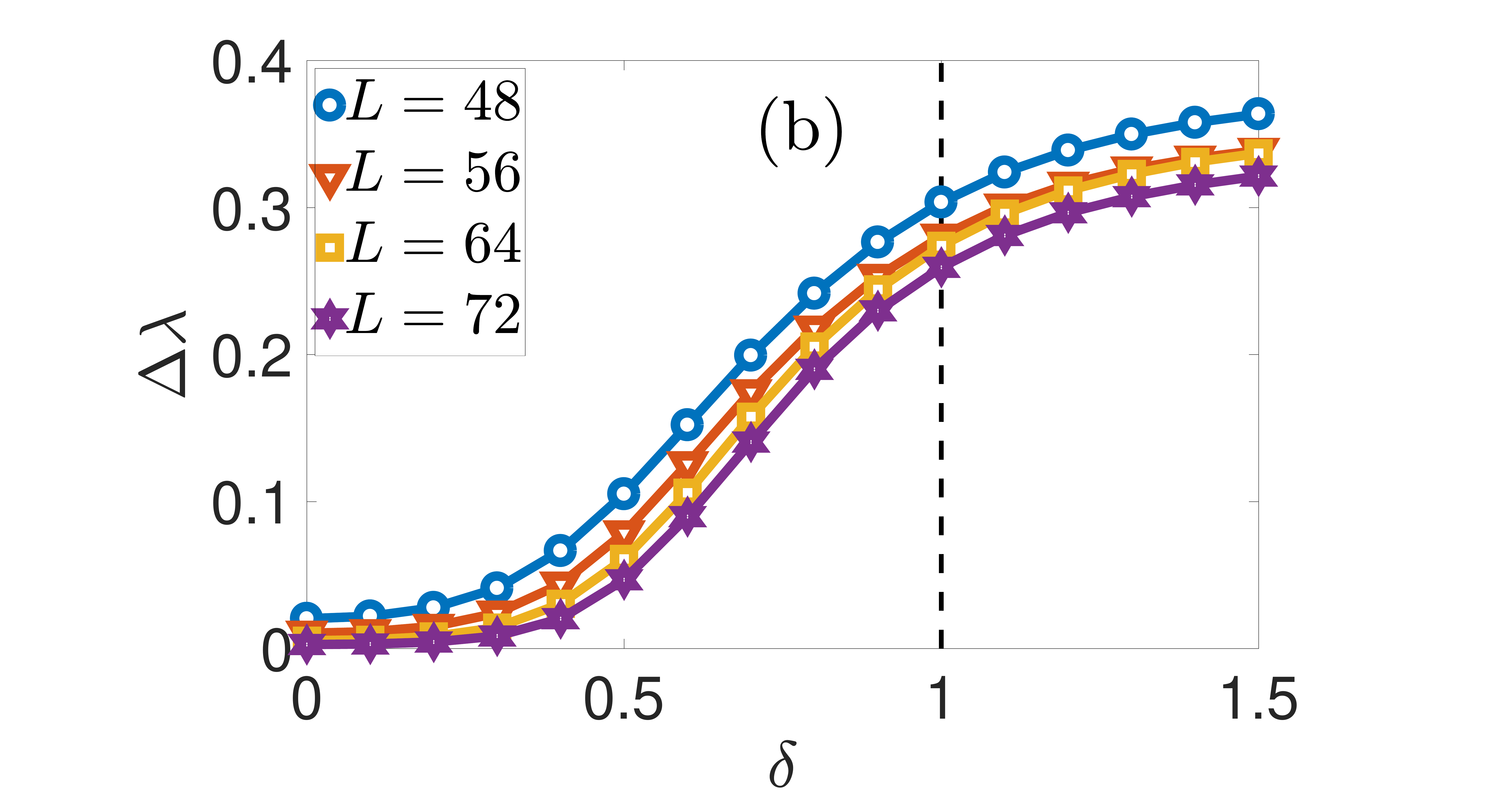}
\includegraphics[trim=6cm 0.45cm 2.5cm 1.5cm, clip=true, width=0.735\columnwidth]{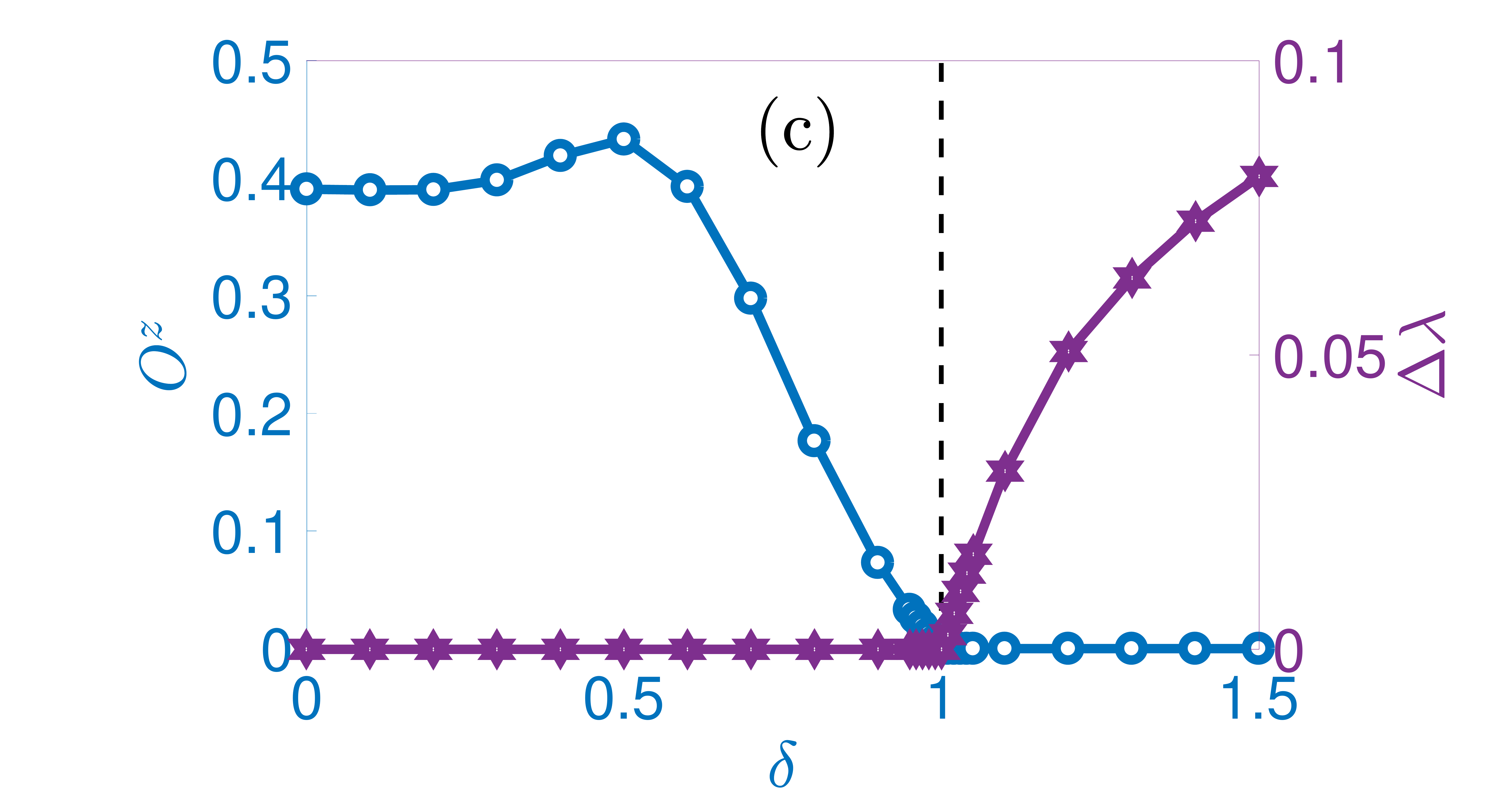}
\caption{(a) Average string order parameter, $O^z$, versus $\delta$ for varying chain length. (b) Average Schmidt gap, $\Delta\lambda$, versus $\delta$; the color coding is the same as in panel (a). (c) Average string order parameter and Schmidt gap extrapolated to infinite length.
In all plots, lines connect points and are a guide for the eye. The dashed vertical line represents the approximate critical value of $\delta$ at which the disorder-induced phase transition takes place.}
\label{fig:SOSG}
\end{figure*}

\subsection{String order and Schmidt gap}
We wish to investigate the disorder induced phase transition from the Haldane gapless phase to the RSP. To this end, we consider two disorder averaged quantities: the string order parameter and the Schmidt gap, introduced in Sec.~\ref{sec:ising}.

The string order parameter is defined as \cite{Nijs89,Ueda08}:
\begin{equation}
O^z=\lim_{|l-r|\to\infty}O^z(l,r),
\end{equation}
where the string correlation function $O^z(l,r)$ is
\begin{equation}
O^z(l,r)=-\left[\left\langle S^z_l\exp\left[i\pi\sum_{k=l+1}^{r-1}S^z_k\right]S^z_r\right\rangle\right]_D,
\end{equation}
and we take the distance ${|l-r|}$ to be approximately $L/2$. The Haldane gapless phase can be detected by the presence of a nonzero string order parameter, since this phase retains long-range correlations from the gapped Haldane phase. On the other hand, the string order vanishes in the RSP, as each spin is only correlated with the spin it is in a singlet with.

Conversely, the disorder-averaged Schmidt gap is expected to be nonzero in the RSP and 0 otherwise. In the RSP there are $3^N$ degenerate ground states, where $N$ is the number of singlets crossing the center of the chain. Therefore, if no singlet crosses the center of the chain, the central two spins are in a product state with corresponding $\Delta\lambda\approx1$. 
For sufficiently large system sizes, there is a nonzero probability of this product state occurring and, in turn, a nonzero disorder-averaged Schmidt gap. However, in the gapped and gapless Haldane phases the ground state is evenly degenerate, corresponding to a Schmidt gap of 0 for all cases of disorder. 

Figure \ref{fig:SOSG}(a) shows the average string order parameter and Fig. \ref{fig:SOSG}(b) shows the average Schmidt gap, for the region of interest ${\delta\in[0,1.5]}$. While the RSP extends for any $\delta>1$, our DMRG calculations become unstable for $\delta>1.5$. Therefore we restrict ourselves to $\delta\le\delta_{max}=1.5$, since we are interested in the transition at ${\delta_c\approx1}$. In Fig. \ref{fig:SOSG}(a) we see a crossing in the string order parameter at ${\delta\approx0.8}$ for the majority of chain lengths, with the crossing occurring at a slightly lower value of $\delta$ for the shorter lengths. It seems reasonable to expect that this crossing would tend towards $\delta_c$ as the chain length increases. While we do not observe a crossing in the Schmidt gap [Fig. \ref{fig:SOSG}(b)], we do see strong evidence of finite-size effects, as it is well known that in the Haldane phase (${\delta=0}$) the Schmidt gap is 0.

Figure \ref{fig:SOSG}(c) shows the results of a finite-size extrapolation to infinite lengths for the two parameters. This finite-size extrapolation is done by implementing a method similar to that of Lajko $et~al$.~\cite{Lajko05}, in which a value for the critical decay exponent $\eta$ is extracted by fitting an algebraic dependence $A/L^\eta$ of the order parameter vs the chain length, where $A$ is a fitting prefactor and we fix ${\delta=\delta_c}$. While the extrapolation for the string order seems to give reasonable results in the regions $\delta\to 0$ and $\delta \sim 1$, we observe a maximum around $\delta\sim 0.5$, which is not observed in the data for fixed lengths. This might be related to the closure of the second energy gap in this region \cite{Lajko05}, leading to a lower quality of the extrapolation for our samples.

It is known \cite{Hyman97,Monthus97,Lajko05} that, for critical disorder, the correlation length diverges as ${\xi\sim(\delta_c-\delta)^{-\nu}}$ with ${\nu=(1+\sqrt{13})/2\approx2.3028}$ and that the string order parameter vanishes as ${O^z\sim(\delta_c-\delta)^{2\beta_{\rm st}}}$ with ${\beta_{\rm st}=2(3-\sqrt{5})/(\sqrt{13}-1)\approx0.5864}$. Therefore, the string order decays with length as ${O^z(L)\sim L^{-\eta_{\rm st}}}$, where ${\eta_{\rm st}=2\beta_{\rm st}/\nu_{\rm st}\approx0.5093}$. However, there is currently no conjecture for the theoretical decay rate of the Schmidt gap, thus we do not have a theoretical prediction for the value of $\eta_{\Delta\lambda}$. {Due to the more conventional construction of the order parameter the Schmidt gap is expected to scale as ${\Delta\lambda\sim(\delta_c-\delta)^{\beta_{\Delta\lambda}}}$, therefore resulting in $\eta_{\Delta\lambda}$ being calculated by $\eta_{\Delta\lambda}=\beta_{\Delta\lambda}/\nu_{\Delta\lambda}$}.
The values we find for the critical exponent $\eta$ obtained for the string order parameter and Schmidt gap are ${\eta_{\rm st}=0.20\pm0.04}$ and ${\eta_{\Delta\lambda}=0.37\pm0.07}$ respectively. Note that the value of $\eta_{\rm st}$ obtained is relatively far from the theoretical value. We expect this discrepancy to be due to the limited sizes of the chains we considered.

\begin{figure}[t] 
\centering
\includegraphics[trim=6.5cm 3.85cm 9cm 2cm, clip=true, width=0.95\columnwidth]{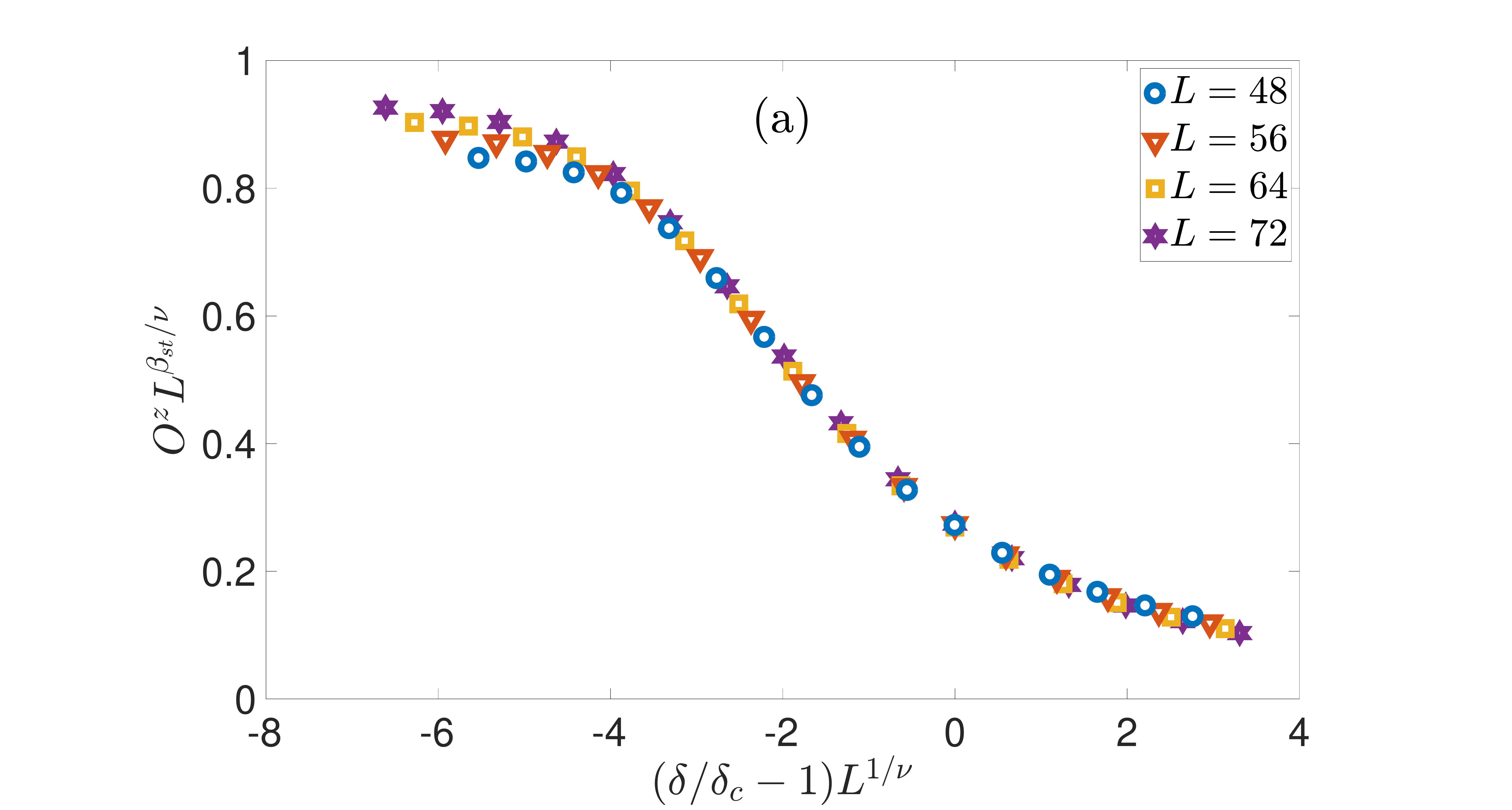}\\[-2.5pt]
\includegraphics[trim=6.5cm 0.45cm 9cm 1cm, clip=true, width=0.95\columnwidth]{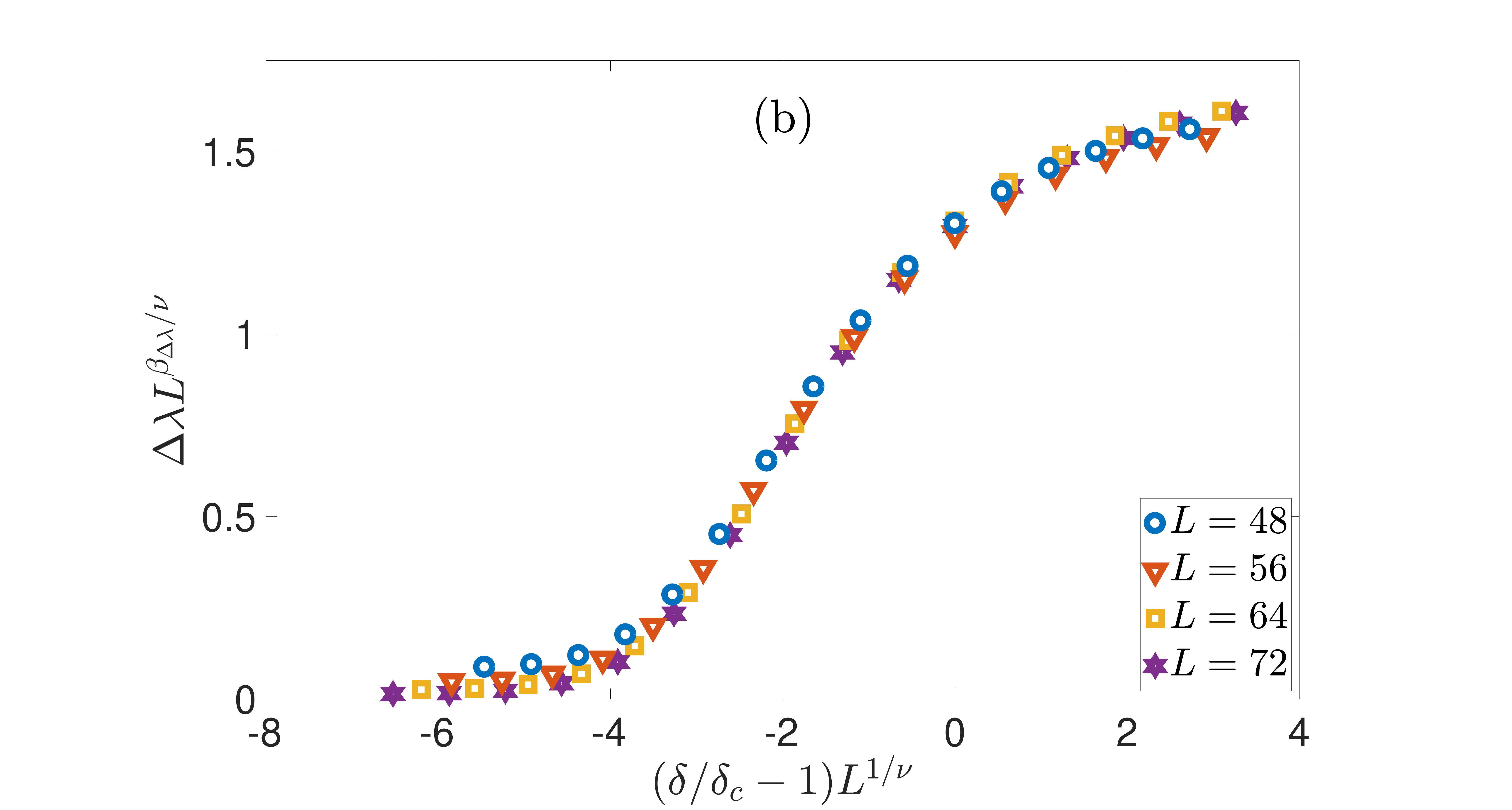}
\caption{(a) Finite size scaling analysis of string order parameter, $O^z$, close to the critical point, $\delta_c$, for chain lengths ${L=48, 56, 64}$ and $72$. (b) Same analysis for the Schmidt gap, $\Delta\lambda$.}
\label{fig:SOSGCollapse}
\end{figure}

We then performed a finite-size scaling analysis \cite{Fisher72} of our results for the string order parameter and the Schmidt gap in order to obtain another approximation of the critical decay exponent using Eq.~\eqref{eq:fss}.
In this work we fix ${\delta_c=1}$ and allow $\nu$ and $\beta_Q$ to vary until the best collapse of the finite-size results is obtained. The string order is known to scale, as above, with ${\beta_Q=2\beta_{\rm st}}$, due to the construction of the order parameter.

Figure \ref{fig:SOSGCollapse} (a) shows the collapse for the string order parameter. The best finite-length collapse was obtained for ${\beta_{\rm st}=0.24\pm0.05}$ and ${\nu_{\rm st}=2.3\pm0.4}$, corresponding to a value of ${\eta_{\rm st}=0.21\pm0.04}$. This, again, is relatively far from the theoretical value of $\eta$ but is in close agreement with the value found previously using the finite-size extrapolation. 
Figure \ref{fig:SOSGCollapse} (b) shows the results when the same finite-size scaling is applied to the Schmidt gap data, with the best collapse obtained at ${\beta_{\Delta\lambda}=0.9\pm0.1}$ and ${\nu_{\Delta\lambda}=2.3\pm0.4}$, corresponding to ${\eta_{\Delta\lambda}=0.38\pm0.08}$. This is significantly closer to the theoretical value of $\eta$ while also being in good agreement with the value found from the extrapolation. In both cases, the critical exponent $\nu$ is found to be very close to the theoretical value, thus validating the numerical simulations. 

\subsection{Entanglement spectrum and entropy}
Finally, we investigate the disorder-averaged ES for the first 12 eigenvalues of the reduced density matrix.
The Haldane phase has a known \cite{PollmannPRB2010,Lepori13} degeneracy sequence of $[2,4,2,4,\dots]$ in the eigenvalues of the reduced density matrix. In the RSP, the ES is dependent on the number of singlets cut at the center of the chain, with eigenvalues $\lambda_1$ to $\lambda_{3^N}$ having a value proportional to $1/3^N$ (with $N$ being the number of spin-1 singlets cut). This leads to an expected eigenvalue degeneracy distribution of $[1,2,6,18,\dots]$ which can also be written as ${[3^0,3^1-3^0,3^2-3^1,3^3-3^2,\dots]}$.

\begin{figure}[t] 
\centering
\includegraphics[trim=6.5cm 0.45cm 8cm 1.5cm, clip=true, width=0.95\columnwidth]{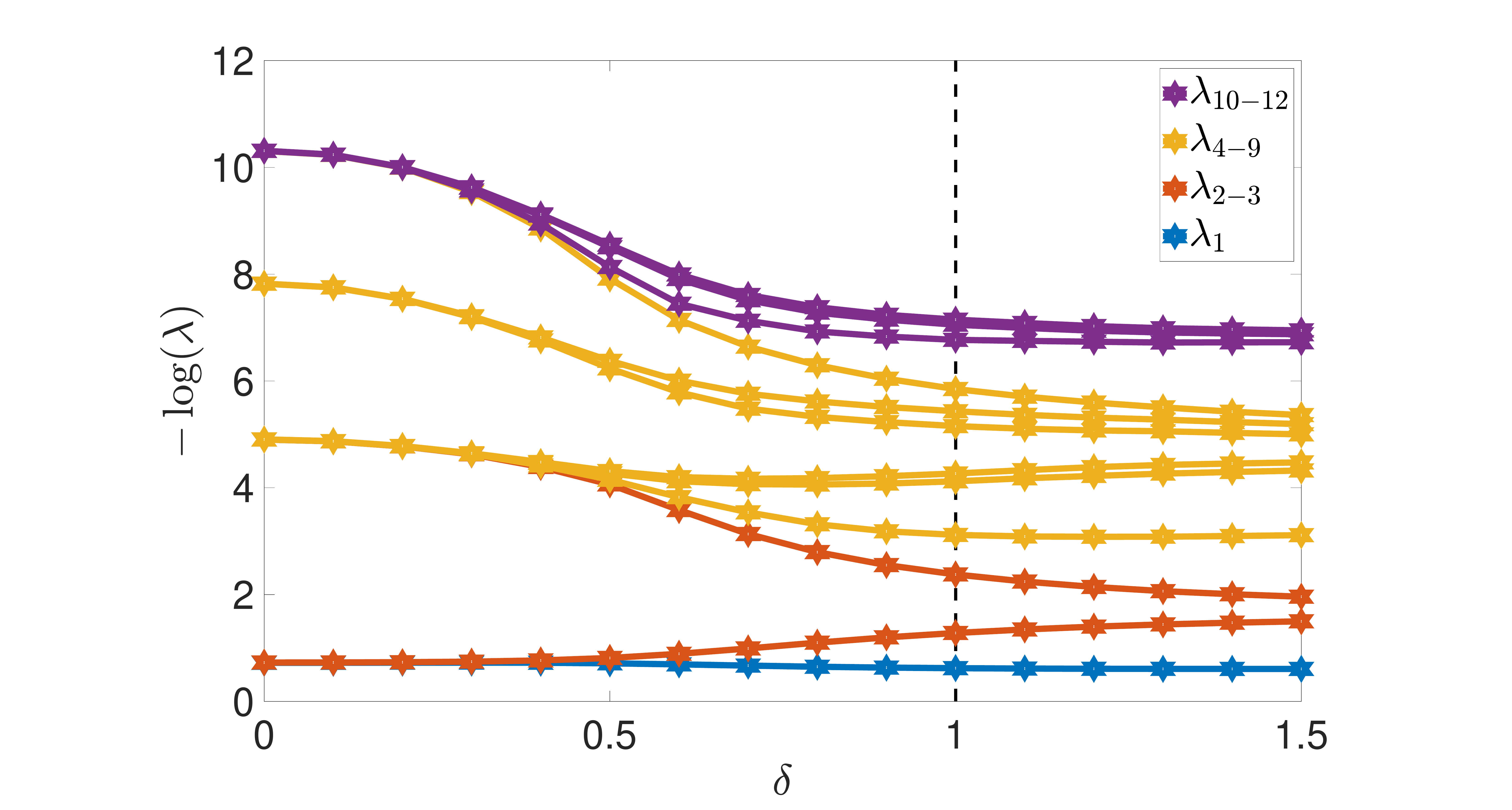}
\caption{
ES, $-\log(\lambda)$, versus $\delta$ for $L=72$. Ascending values of $\lambda_i$ with $i=1,2,\dots12$ as the magnitude of $-\log(\lambda)$ increases. Colors correspond to the expected eigenvalue grouping in the RSP.  The dashed vertical line, as in Fig.~\ref{fig:SOSG}, represents the approximate critical value, $\delta_c$.}
\label{fig:EntSpec}
\end{figure}

Figure \ref{fig:EntSpec} shows the disorder averaged ES for the first 12 eigenvalues of a chain of 72 spins. For this fixed length, the the eigenvalues separate significantly for ${\delta>0.4}$. We expect that in the thermodynamic limit this separation would occur close to $\delta_c$. We can observe quite clearly the transition between the Haldane phase and the RSP in the structure of the ES, shown in Fig. \ref{fig:EntSpec}. In particular, we expect that for larger values of $\delta$, where all contributions but those of singlets are almost eradicated, $\lambda_4$ will group closer with the eigenvalues $\lambda_5$ and $\lambda_6$.

Particularly interesting is also the probability distribution $P$ of the entanglement entropy, as it is directly related to the distribution of the eigenvalues.
We calculate the von Neumann entropy:
\begin{equation}
\label{eq:EE}
E=-\mathrm{Tr}\rho_\ell\log_2\rho_\ell
\end{equation}
of the reduced density matrix
\begin{equation}
\rho_\ell=\mathrm{Tr}_{L-\ell}|\psi_G\rangle\langle\psi_G|,
\end{equation}
where $|\psi_G\rangle$ is the ground state of Hamiltonian \eqref{Hamiltonian} and $\ell=L/2$.

\begin{figure}[t] 
\centering
\includegraphics[trim=4cm 0.45cm 12cm 1.5cm, clip=true, width=0.475\columnwidth]{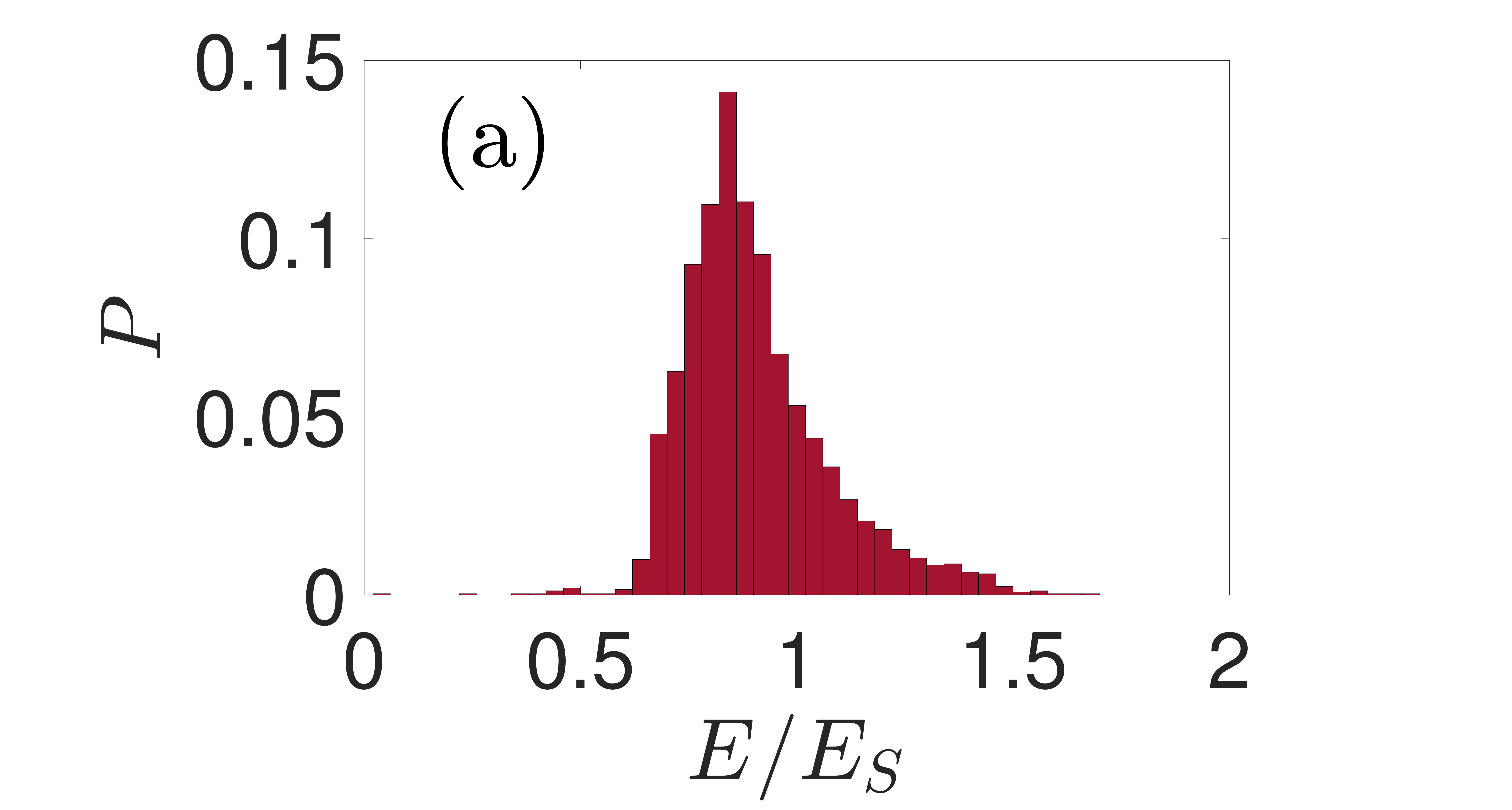}
\includegraphics[trim=8cm 0.45cm 8cm 1.5cm, clip=true, width=0.475\columnwidth]{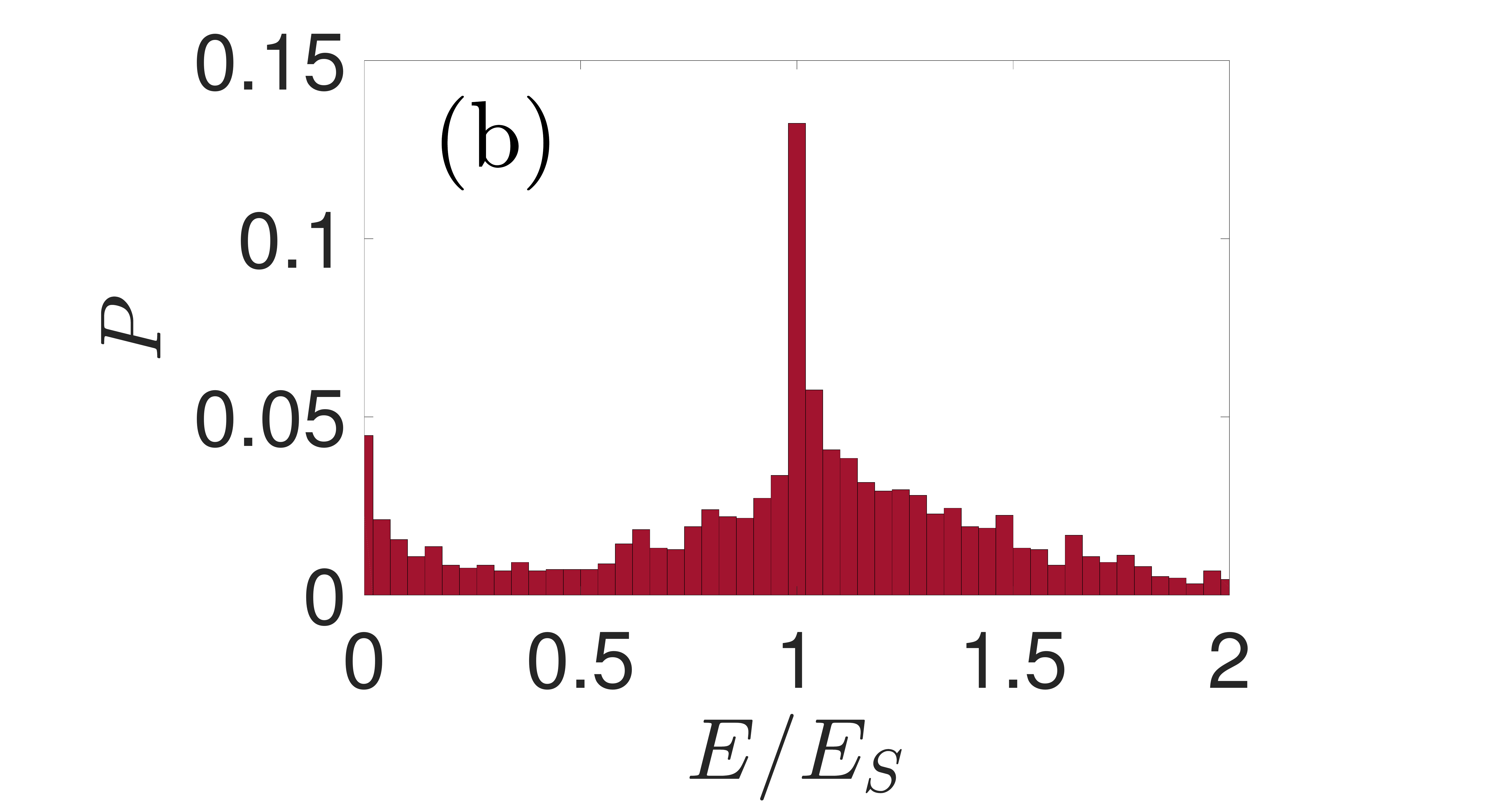}
\caption{Probability distribution, $P$, of the von Neumann entropy, $E$, for strength of disorder (a) ${\delta=0.5}$ and (b) ${\delta=1.5}$ for a spin chain of length 56.
}
\label{fig:VNDist}
\end{figure}

Figure \ref{fig:VNDist} shows the probability distribution of the von Neumann entropy for two values of disorder, one corresponding to the Griffiths phase and one to the RSP. The distribution is plotted such that the horizontal axis represents the ratio $E/E_S$, where ${E_S=\log_2(3)\simeq 1.585}$ is the entanglement of a spin-1 singlet. Therefore, peaks at integer values represent an integer number of singlets crossing the center of the chain. A significant change in the distribution of entanglement is seen as we move from the Griffiths phase to the RSP. Specifically, the distribution becomes much broader in the RSP but at the same time we see a dominance of one entanglement value (and thus ES), which is unseen in the Griffiths phase. Our ability to only investigate smaller values of $\delta$ explains the relatively rare occurrence of zero ${(\sim5\%)}$ and more than one ${(\sim0.5\%)}$ spin-1 singlets being cut. As such, we fully expect these peaks to become more prominent for larger disorders and for larger lengths.

It is well understood \cite{Laflorencie05,Binosi07} that, further in the RSP for large disorder, the smearing between contributions to the von Neumann entropy from singlets decreases, and the same would be seen in the distribution of the eigenvalues. We assume that, for strong enough disorder, the ES will depend exclusively on the number of singlets being cut, and thus the disorder-averaged spectrum will depend on the probability of cutting a number of singlets $N$, with this probability varying as the disorder increases.

\section{Conclusions}
In summary, we have numerically investigated the entanglement spectrum of the ground state of random spin-1/2 and spin-1 chains. The structure and degeneracy of the low-lying levels of the entanglement spectrum reveal the emergence of a quantum phase transition even in these disordered models.
Remarkably, even for the two inequivalent random models we studied, the Schmidt gap detects the corresponding critical points and scales with universal critical exponents. 
These results reinforce the role of the Schmidt gap as a useful probe in quantum critical phenomena and open the way to possible extensions to dynamics in the presence of disorder and noise. 

\acknowledgements
We thank E. Canovi, E. Ercolessi, and L. Taddia for fruitful discussions. This research was supported by the Ontario Trillium Foundation, the Perimeter Institute for Theoretical Physics, through Industry Canada, and the Province of Ontario, through the Ontario Ministry of Research, Innovation and Science.

\bibliographystyle{apsrev4-1}
\bibliography{biblio}
\end{document}